\documentclass[12pt,preprint]{aastex}

\newcommand{\glimpse} {\mbox{\it GLIMPSE~}}

\newcommand{\irac}{\mbox {\it IRAC~}}
\renewcommand{\deg}{$^\circ$}

\def\etal{\hbox{\it \etal}}

\slugcomment{Accepted for publication in Astrophysical Journal Letters}

\shorttitle{\glimpse Stellar Structure  }
\shortauthors{Benjamin {\it et al.} }

\begin{document}


\newlength{\hugepicsize}
\setlength{\hugepicsize}{5.5in}
\newlength{\bigpicsize}
\setlength{\bigpicsize}{4.5in}
\newlength{\smpicsize}
\setlength{\smpicsize}{3.5in}
\newlength{\tinypicsize}
\setlength{\tinypicsize}{2.5in}
\newlength{\tinierpicsize}
\setlength{\tinierpicsize}{2.0in}

\title{ First GLIMPSE Results on the Stellar Structure of the Galaxy}


\author{ 
Benjamin, R.~A.\altaffilmark{1},
Churchwell, E.\altaffilmark{2},
Babler, B.~L.\altaffilmark{2},
Indebetouw, R.\altaffilmark{3},
Meade, M.~R.\altaffilmark{2},
Whitney, B.~A.\altaffilmark{4},
Watson, C.\altaffilmark{5},
Wolfire, M.~G.\altaffilmark{6},
Wolff, M.~J.\altaffilmark{4}, 
Ignace, R.\altaffilmark{7}
Bania, T.~M.\altaffilmark{8},
Bracker, S. \altaffilmark{2},
Clemens, D.~P.\altaffilmark{8},
Chomiuk, L.\altaffilmark{2},
Cohen, M.\altaffilmark{9},
Dickey, J.~M.\altaffilmark{10}, 
Jackson, J.~M.\altaffilmark{8},
Kobulnicky, H.~A.\altaffilmark{11}, 
Mercer, E.~P.\altaffilmark{8}, 
Mathis, J.~S.\altaffilmark{2},
Stolovy, S.~R.\altaffilmark{12},
Uzpen, B.\altaffilmark{11}
}

\altaffiltext{1}{University of Wisconsin-Whitewater, Dept. of Physics,
800 W. Main St., Whitewater, WI 53190}

\altaffiltext{2}{University of Wisconsin-Madison, Dept. of Astronomy,
475 N. Charter St., Madison, WI 53706}

\altaffiltext{3}{University of Virginia, Dept. of Astronomy,  P.O. Box 3818, 
Charlottesville, VA 22903-0818}

\altaffiltext{4}{University of Colorado, Space Science Institute,
1540 30th St., Suite 23, Boulder, CO 80303-1012}

\altaffiltext{5}{Manchester College, Physics Dept., Box 117, 
North Manchester, IN 46962}

\altaffiltext{6}{University of Maryland, Astronomy Dept., College Park,
MD 20742-2421}

\altaffiltext{7}{East Tennessee State University, Dept. of Physics, Astronomy \& Geology,  PO Box 70652,  Johnson City, TN 37614}

\altaffiltext{8}{Boston University, Institute for Astrophysical
Research, 725 Commonwealth Ave., Boston, MA 02215}

\altaffiltext{9}{University of California-Berkeley, Radio Astronomy Lab,
601 Campbell Hall, Berkeley, CA 94720}

\altaffiltext{10}{University of Tasmania, School of Mathematics and Physics,
 Private Bag 21, Hobart, Tasmania, Australia 7001}

\altaffiltext{11}{University of Wyoming, Dept. of Physics \& Astronomy,
PO Box 3905, Laramie, WY 82072}

\altaffiltext{12}{Caltech, Spitzer Science Center, MS 314-6, Pasadena,
CA 91125}

\begin{abstract}
The \glimpse (Galactic Legacy Mid-Plane Survey Extraordinaire) Point Source Catalog of $\sim$ 30 million mid-infrared sources towards  the inner Galaxy,  $10^{\circ} \le |l| \le 65^{\circ}$ and $|b| \le 1^{\circ}$, was used to determine the distribution of stars in Galactic longitude, $l$, latitude, $b$, and apparent magnitude, $m$.  The counts  versus longitude can be approximated by the modified Bessel function  $N=N_{0}(l/l_{0})K_{1}(l/l_{0})$, where $l_{0}$ is insensitive to limiting magnitude, band choice, and side  of Galactic center:   $l_{0}= 17^{\circ}-30^{\circ}$ with a best fit value in the the  4.5$\mu$m band of $l_{0}=24^{\circ} \pm 4^{\circ}$.  Modeling the source distribution as an exponential disk yields a radial scale length of $H_{*}= 3.9 \pm 0.6$ kpc.  

There is a pronounced north-south asymmetry in source counts for $|l|^{<}_{\sim} 30^{\circ}$, with $\sim$25\%  more stars in the north.  For $l=10-30^{\circ}$, there is a strong enhancement of stars of $m= 11.5-13.5$ mag. A linear bar passing through the Galactic center with half-length $R_{bar}=4.4 \pm 0.5$ kpc, tilted by  $\phi=44^{\circ} \pm 10^{\circ}$ to the Sun-Galactic Center line, provides the simplest interpretation of this data.  We examine the possibility that enhanced source counts at  $l=26-28^{\circ}$, $31.5-34^{\circ}$, and $306-309^{\circ}$ are related to Galactic spiral structure. Total source counts are depressed  in regions where the counts of red objects ($m_{K}-m_{[8.0]} >3$) peak. In these areas, the counts are reduced by extinction due to molecular gas and/or high diffuse backgrounds associated with star formation. 
\end{abstract}


\keywords{infrared:general -- infrared:ISM -- infrared:stars -- survey -- stars:general -- ISM:general -- Galaxy:structure -- Galaxy:stellar content }

\section{\glimpse and the stellar structure of the Galaxy}

Since the pioneering work of Kapteyn (1922),  star counts have been an important avenue  in exploring the stellar structure of the Galaxy.   However, progress on understanding the stellar content of the inner Galactic disk has been impeded by dust obscuration of visible light from stars. With the advent of infrared detectors and surveys of increasingly high sensitivity and resolution, significant progress has been made, leading, for example, to strong evidence for a Galactic bar (reviewed by Gerhard 2002).  



\glimpse, a Spitzer Legacy Science Program, has imaged the Galactic plane from Galactic longitude\footnote{$|l| \le 180^{\circ}$ is the angular distance from Galactic center.} $|l|$ = 10\deg\ to 65\deg\  and Galactic latitude  $|b| \le$ 1\deg\ at 3.6, 4.5, 5.8, and 8.0 $\mu$m using the \irac (Infrared Array Camera, Fazio {\it et al.} 2004) on the {\it Spitzer Space Telescope} (Werner {\it et al.} 2004).  Angular resolutions range  from 1.4$\arcsec$ at 3.6 $\mu$m to 1.9$\arcsec$ at 8 $\mu$m. We present the mid-infrared distribution of stars in the Galaxy as determined by  \glimpse. Properties of the \glimpse Point Source Catalog and Archive are summarized in Table 1. Observing strategy, details of  source selection, quality flags, and photometric accuracy are provided in the \glimpse Science Data Products document (Meade {\it et al.} 2005), the \glimpse Quality Assurance Document (Churchwell {\it et al.} 2005), and Benjamin {\it et al.} (2003).

We have  compared the results obtained for all four \irac bands,  explored the effect of using the Archive  instead of the Catalog, explored the sensitivity of our results to our faint magnitude limit, divided the survey region into three latitude strips of $\Delta b=0.3^{\circ}$ and considered each strip separately, and experimented with the effect of removing sources with various quality flags. The parameters of all the principal features discussed below remained unchanged. 

\section{ \glimpse Counts as a Function of Longitude and Latitude}

The Catalog and Archive sources were binned into data cubes of 0.1\deg (longitude) $\times$ 0.1\deg (latitude) $\times$ 0.1 magnitude for all four \irac bands. The average number of sources  in each 0.1\deg $\times$ 0.1\deg\ bin is $\sim$1,400 for the Catalog and $\sim$2,200 for the Archive.  Figure 1 shows the average number of  4.5$\mu$m sources per square degree in the magnitude range $m=6.5$ to $m=12.5$.  The curves indicate the source-count weighted Galactic latitude distribution, $b_{cen}(l)=\frac{\Sigma_{b} N(l,b)b}{\Sigma_{b} N(l,b)}$ , which ranges from $b_{cen}=-0.05^{\circ}$ to $+0.05^{\circ}$.  As $|l|$ increases from 10\deg\ to 65\deg, the counts decrease by a factor of $\sim$ 12, but there is a pronounced asymmetry between the north and the south in the inner Galaxy.  From 10\deg-22\deg\ there are $\sim$ 25\%  more sources than in the equivalent southern longitude range.  There are also apparent enhancements in the counts for $l=26^{\circ}-28^{\circ}$, $31.5^{\circ}-34^{\circ}$, and $306^{\circ}-309^{\circ}$. Figure 1 also shows contours of the surface density of $\sim 40,000$ unusually red ($m_{K}-m_{[8.0]} \ge 3.0$) sources using the  \glimpse and {\it 2MASS} Catalogs (Cutri et al. 2001).  These objects are strongly anti-correlated with the total \glimpse counts. The regions of depressed counts coincide with high column densities of $^{13}CO$ (Simon {\it et al.} 2001; Shah 2005, priv. comm.) or thermal radio continuum emission (i.e. high mass star formation). 

Figure 2 shows that the number of sources per square degree brighter than magnitude $m$ is well-fit by  the first order modified Bessel function of the second kind, $N({\rm band},m, |l|)=N_{0} (|l|/l_{0})K_{1}(|l|/l_{0})$, which is expected for an exponential disk population (Merrifield, priv. comm.). Our ``best-fit'' curve excludes data  interior to $|l|=35^{\circ}$, from  $l=306.5-309^{\circ}$ (enhancement) and $l=301.5-306.5^{\circ}$ (missing data). The inner boundary was varied from $|l|=25^{\circ}$ to $|l|=45^{\circ}$ to estimate systematic uncertainties due to this choice. The range for all the \irac bands is $l_{0}= 17-30^{\circ}$. The best fit in the 4.5$\mu$m band is $l_{0}=24.2 \pm 0.3^{\circ} {\rm (random)} \pm 3.6^{\circ}$ {\rm ($2\sigma$ systematic)}; the slope and  level of the fits for the north and south are in marginal agreement.  Comparison of this fit to numerical models of a disk population of M and K giants, which are assumed to dominate the source counts (Wainscoat {\it et al.} 1992), yields an exponential scale length of $H_{*}=3.9 \pm 0.6$ kpc.   

For longitudes greater than $|l|=30^{\circ}$ and excluding the enhancements mentioned above, the fitting functions for the 4.5$\mu$m band predict the observed source densities to within 20\% for a faint magnitude cutoff of 8 and 10 mag, and to within 10\% for faint magnitude cutoffs of 12 and 14 magnitudes. Interior to $|l|=22^{\circ}$, the fits overpredict the observed counts by as much as 30\% in both the northern and southern plane.  This is not due to confusion as the drop is independent of the magnitude cut; even at faint magnitudes, we are not affected by confusion (Churchwell {\it et al.} 2005). 

\section{\glimpse Counts as a Function of Magnitude}

Figure 3 shows the number of sources as a function of magnitude for three north/south pairs of directions.  Figures 3 and 4 were used to determine the  ``effective'' sensitivity limit, $m_{sens}$, as a function of longitude given in Table 1.  The curves shown in Figure 3 can be fit by  $\log_{10}n(m)=am + b$. Since  $m=-2.5 \log_{10}(S/S_{0})$, where S is the flux density at magnitude $m$ and $S_{0}$ is the flux density for zero magnitude, we can write $N(S)=N_{0}  \left( \frac{S}{S_{0}} \right) ^{-\alpha}$, where $N_{0}=(2.5 /S_{0}\ln 10)10^{b}~ {\rm sources~deg^{-2}~Jy^{-1}}$ and $\alpha=2.5 a+1$ .   The average  $\bar{\alpha}$ over the magnitude range $8<m<11$ lies in the range $1.95$ to $1.83$ for the entire survey area.

Figure 3 shows a ``hump'' at magnitudes fainter than $m=12$ in the inner Galaxy towards the north but not the south. This hump is seen in both the Catalog and Archive data in the 3.6$\mu$m and 4.5$\mu$m bands.  It is not detected in the 5.8$\mu$m and 8.0$\mu$m bands due to lower sensitivity. To study this further, in Figure 4 we plotted the power-law slope, $\alpha=2.5\frac{d(\log n)}{dm}+1$ (with $\Delta m=0.1$) of the source distribution as a function of Galactic longitude and magnitude. The hump persists over the range $l=$10-22\deg\  and  $l=$24-29\deg .  It seems likely that the gap at $l \sim$ 23\deg\ is caused by high extinction or high diffuse flux. The  central magnitude of the hump at $l=10^{\circ}$ is $m_{h}=12.5\pm0.1$ mag with a dispersion of $\sigma_{h}=0.25\pm0.10$ mag; the amplitude at the peak is  20\% above the power-law fit.  The shift of the central magnitude of the hump  to brighter magnitudes with increasing Galactic longitude is given by $dm_{h}/dl=-0.025\pm0.005~{\rm mag~deg^{-1}}$. 

\section{Interpretation and Comparison with Previous Results} 

We now summarize the principal features we have found and possible implications for our understanding of Galactic structure. We assume the distance to the Galactic center is $R_{0}=8.5$ kpc. 

 {\it Exponential Disk:}  For $|l| ^{>}_{\sim} 22^{\circ}$, the counts in all four \irac bands can be fit by $N({\rm band},m, |l|)=N_{0} (|l|/l_{0})K_{1}(|l|/l_{0})$. For the 4.5 $\mu$m band, the best fit  (excluding regions of obvious enhancement) yields  $l_{0}=24.2 \pm 3.6^{\circ} {\rm (systematic)}$. For all four \irac bands and for several magnitude cutoffs, we find that $l_{0}=17^{\circ}-30^{\circ}$. This can be compared with our results obtained by binning the {\it 2MASS}  K band sources in an identical fashion which yields $l_{sc} > 40^{\circ}$. Presumably this larger number is due to higher extinction in the K band.  Converting this angular scale length to a radial scale length requires modeling of the spatial distribution of stars, the infrared luminosity function, and extinction (Cohen 1993). Preliminary modeling indicates that the majority of our sources are M giants and late K giants, and yields a radial scale length for the exponential disk of $H_{*}=3.9 \pm 0.6$ kpc. 
Interior to $|l|=22^{\circ}$ there are $\sim$ 20\% fewer sources than predicted by this fit, which may indicate a stellar ``hole'' (or more accurately, deficit)  interior to Galactocentric radius, $R{h}=3.2\pm0.3~{\rm kpc}$.  This is comparable to $R_{h}=2.9-3.3$ kpc derived from {\it COBE/DIRBE} data (Freudenreich 1998). 


 {\it Galactic Bar:}  We see a strong  north/south asymmetry in counts for $|l| ^{<}_{\sim} 30^{\circ}$, with $\sim$25\% more sources in the north and a distinct hump in the number of stars at apparent magnitude $m \sim 12.5$ over the longitude range $l=10-30^{\circ}$. Both of these facts suggest that we have detected the Galactic bar. Estimates for the bar orientation (relative to the line connecting the Sun to the Galactic center) typically range from $\phi=15^{\circ}-35^{\circ}$ (Gerhard 2002) . If  the ``hump'' seen in Figure 3 and 4 is due to stars in the bar with a well-defined peak at constant absolute magnitude, we can estimate the bar orientation and length. In Figure 4, we have plotted the magnitude and longitude expected for a linear bar in the plane of the Galaxy passing through the Galactic center  with a foreground extinction $a_{[4.5]}(d)=0.05~ {\rm mag~kpc^{-1}}$, using the fiducial $a_{V}(d)= 1~{\rm mag~kpc^{-1}}$(Mihalas \& Binney 1981), $A_{V}/A_{K}=8.8-7.5$ (Cardelli, Clayton, \& Mathis 1989) and $A_{[4.5]}/A_{K}=0.43$ (Indebetouw {\it et al.} 2005). The best fit to the data yields  $\phi=44 \pm 10^{\circ}$, $R_{bar}=4.4 \pm 0.5$ kpc, and $M_{[4.5]}=-2.15\pm 0.2$ mag; the effect of increasing extinction is to increase $\phi$ and $M$. Fitting the data with zero extinction yields $M_{[4.5]}=-1.8$ mag. The requirement that $M_{[4.5]}=-2.15$ mag to match the data suggests that the hump is due to K2-3 giants (Cohen 1993; Hammersley {\it et al.} 2000).  Figure 4 shows that the  predicted $m_{hump}$ for the model bar falls below our sensitivity limit at $l=350^{\circ}$, but the observed increase in counts seen in Figure 1 from  $l=$345\deg\ to $l=$350\deg\ is consistent with our derived bar length. Although a bar seems to be the simplest interpretation of these results, other configurations should be explored.



 {\it Stellar Enhancements/Spiral Arms(?):} Outside $l=22^{\circ}$, we find three regions where counts are 20\% higher than the exponential fitting functions. These are at $l=26-28^{\circ}$, $31.5-34^{\circ}$,  and $306-309^{\circ}$. It is tempting to associate these features with spiral arm tangencies. The presence of the ``hump'' in the $l=26-29.5^{\circ}$ region argues that it is a continuation of the same structure observed from $l=10-20^{\circ}$, complicated by the presence of extinction. Unresolved questions regarding extinction also affect interpretation of the $31-34^{\circ}$  feature.  The third feature could be identified with a previously known spiral arm tangency (Centaurus, $l \sim 309^{\circ}$).  No plausible enhancement is star counts is seen towards the  Sagittarius spiral arm tangency at $l \sim 49^{\circ}$.     


 {\it Extinction/Diffuse Background:}  There is a clear anticorrelation between the \glimpse\ Catalog counts and the counts of sources with $m_{K}-m_{[8.0]}>3.0$  in certain regions (shown in Figure 1). Counts can be depressed by either increased extinction or high diffuse background, both of which are associated with star forming regions and molecular clouds. It is clear that the effects of extinction and diffuse emission on the source counts deserves future scrutiny. For this reason, the parameters of the structures discussed here should be considered preliminary. 

Many of the principal structures seen in this work have been detected or hinted at in previous infrared investigations. What differentiates the \glimpse results from these previous works is the uniform and complete sampling of the Galactic plane at $ ^{<}_{\sim} 2$ arsec resolution,  the significantly reduced extinction in the mid-infrared, and the very large sample of \glimpse sources.  Even without any color-selection of our sources, we find that we are able to detect and confirm several fundamental Galactic structures. 

\acknowledgments We thank the referee,  Stephan Jansen, Mike Merrifield, Paul Rybski, and Ronak Shah for help and suggestions, and acknowledge support for this program through NASA contracts 1224653 (UW-W), 1224653 (UW-M), 1225025 (BU), 1224681 (UMd), 1224988 (SSI), 1259516 (UCB), 1253153 (UMn), 11253604 (UWy) by the Jet Propulsion Laboratory, Caltech under NASA contract 1407. We also gratefully acknowledge use of data products from the Two Micron All Sky Survey.

\begin{figure}
\plotone{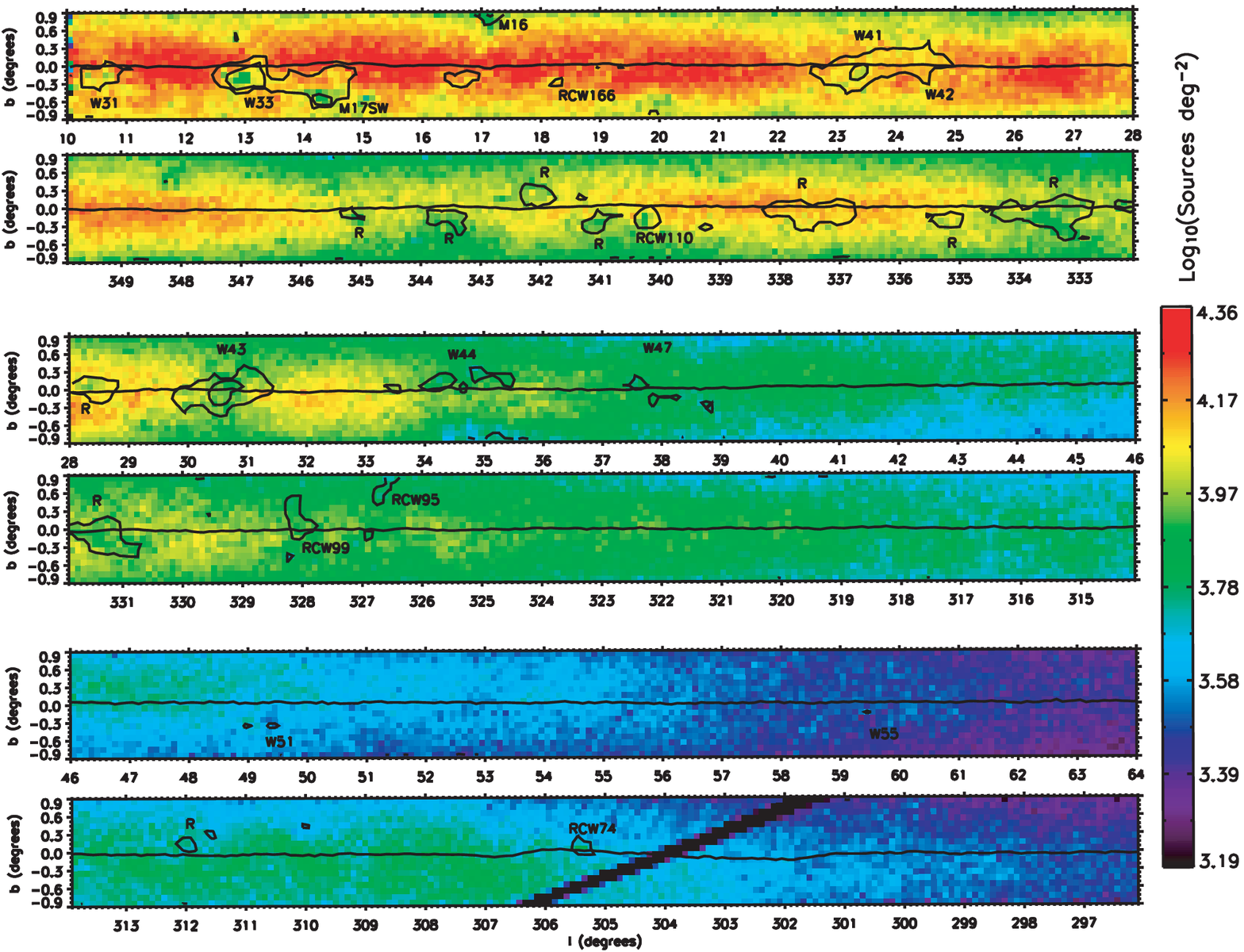}
\caption{${\rm Log_{10}(sources~deg^{-2})}$ of 4.5 $\mu$m band sources from the \glimpse Point Source Catalog in the magnitude range $6.5 < m < 12.5$ as a function of Galactic latitude and longitude. Sources have been binned by $0.1^{\circ}$ (longitude) $\times$ $0.1^{\circ}$(latitude). Notable features include a decline in the number of sources with distance from Galactic center and a  pronounced north-south asymmetry in the inner Galaxy between the north($l=10-22^{\circ}$) and south ($l=350-338^{\circ}$). The number of sources is nearly symmetric about the Galactic midplane; the solid line shows the position of the source-count weighted average latitude. The two contours, ${\rm log_{10}(sources~deg^{-2})}=2.8,~3.3$ show the surface density of  red  ($m_{K}-m_{[8.0]} > 3$) sources;  note the anti-correlation with the total counts. These directions are labeled with corresponding star forming regions or {\bf R} for strong radio continuum (Altenhoff {\it et al.} 1970; Haynes, Caswell, \& Simons 1978). The dark stripe from $l=301-306^{\circ}$ is a region of missing data.}
\end{figure}

\begin{figure}
\epsscale{.80}
\plotone{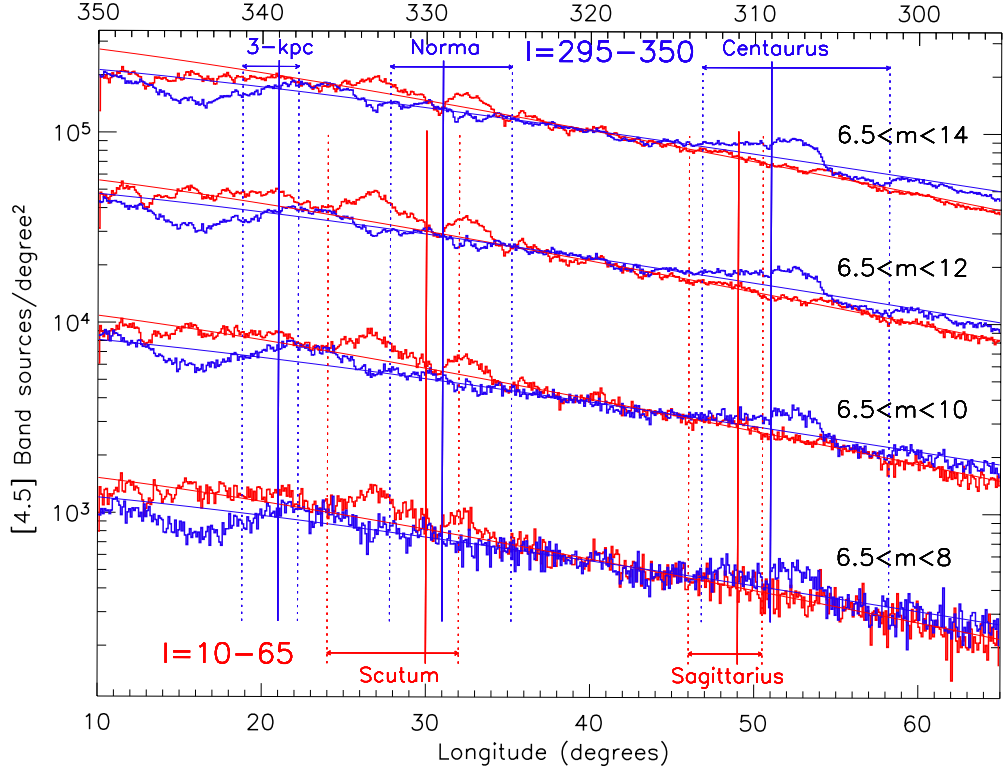}
\caption{Average number of sources per square degree detected by \glimpse as a function of longitude in both the northern Galactic plane (red curves, coordinates on lower axis) and southern Galactic plane (blue curves, coordinates on upper axis). Average source densities were obtained by binning the stars into strips of 0.1 degree in longitude and 1.8 degrees in latitude. Fits to both curves (see text) over the range $l =35^{\circ} - 65^{\circ}$ (excluding regions of clear enhancements)  are also shown. The four sets of curves show the effect of changing the faint mangitude cutoff. For each drop in magnitude, the number of sources increases by a factor of 3.5 (at 9 mag) to 2.3 (at 13 mag).  The vertical lines indicate the estimated directions of spiral arms using different tracers (Englmaier  \& Gerhard 1999). The solid line shows their adopted values;  the dotted lines indicate the range of published estimates. }
\end{figure}

\begin{figure}
\epsscale{1.0}
\plotone{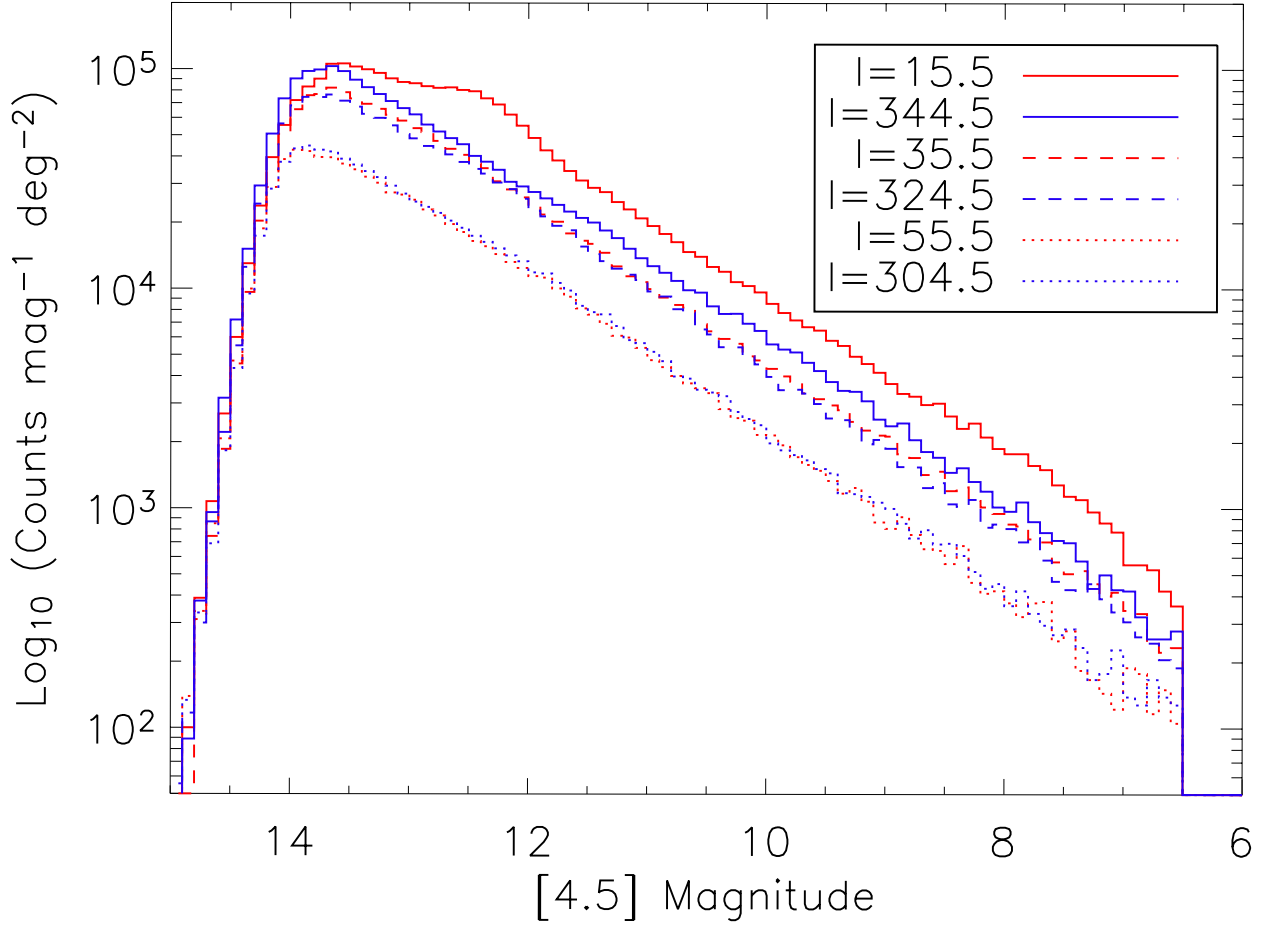} 
\caption{Number of sources from the \glimpse Point Source Catalog as a function of magnitude for three northern and three southern directions. The outer galaxy ($l=55.5^{\circ}$/$l=304.5^{\circ}$) and middle galaxy ($l=35.5^{\circ}$/$l=324.5^{\circ}$) curves have approximately the same amplitude and slopes. 
The inner  galaxy ($l=15.5^{\circ}$/$l=324.5^{\circ}$) shows a significant north/south asymmetry; the northern direction also shows a bump in source counts at a magnitude of $m \sim 12.2$. For all directions shown, the number of sources has been averaged over a  1.0 (longitude) $\times$ 1.8 (latitude) region. The Catalog is truncated at 6.5 mag due to detector nonlinearity for sources brighter than this limit. }
\end{figure}

\begin{figure}
\epsscale{.65}
\plotone{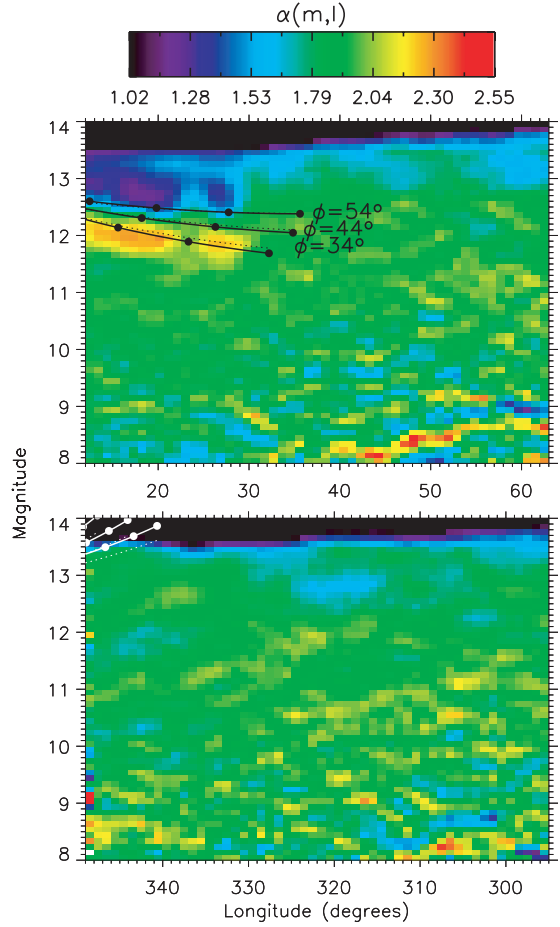}
\caption{ Power-law exponent of counts as a function of flux density  plotted as a function of apparent magnitude and Galactic longitude. The position of the $\sim12$ mag hump seen towards $l=15.5^{\circ}$ in Figure 3, is seen here to vary consistently in both longitude and magnitude.  The locus of magnitude and longitude  of a model bar consisting of stars of absolute magnitude $M_{[4.5]}=-2.15$, foreground extinction $a_{[4.5]}(r)=0.05~{\rm mag~kpc^{-1}}$ and three different position angles $\phi$, are shown in black in the upper panel and in white in the lower panel.  The dots indicate $R=$3, 4, and 5 kpc points along the bar. The dotted lines show the same position angles for zero extinction, with $M_{[4.5]}=-1.8$.  }
\end{figure}





\begin{deluxetable}{ccccccccccc}
\tabletypesize{\small}
\tablecaption{GLIMPSE Catalog/Archive Source Information} 
\tablehead{ 
\colhead{IRAC}   &  \colhead{$\lambda$\tablenotemark{a}}  & \colhead{$S_{0}$\tablenotemark{a}}  & 
\colhead{$\frac{A_{[\lambda]}}{A_{K}}$\tablenotemark{b}}   &  \colhead{$m_{sel}$\tablenotemark{c}} &  \colhead{$m_{br}$\tablenotemark{c}}  & \colhead{$m_{sens}$\tablenotemark{d}} & \multicolumn{2}{c}{Catalog Sources\tablenotemark{e}} & \multicolumn{2}{c}{Archive Sources\tablenotemark{e} }   \\ 
\colhead{Band }  &  \colhead{($\mu$m)}   & \colhead{(Jy)} & \colhead{ }     &
 \colhead{mag} & \colhead{mag} & \colhead{mag} & \colhead{North}  & \colhead{South\tablenotemark{f}}  & \colhead{North}  &\colhead{South}\tablenotemark{f} }
\startdata
1   & 3.55 & 277.5 & $0.56 \pm 0.06$  & 14.2 & 7.0 & 13.3$-$13.6      & 14.775  & 14.255   &  21.420 & 22.044     \\ 
2   & 4.49 & 179.5 & $0.43 \pm 0.08$  & 14.1 & 6.5 & 13.3$-$13.6      & 14.768  & 14.250   &  19.797 & 19.423     \\ 
3   & 5.66 & 116.5 & $0.43 \pm 0.10$  & 11.9 & 4.0 & 11.7$-$12.3      &  5.768  &  5.291   &   6.095 &  5.594     \\ 
4   & 7.84 & 63.13 & $0.43 \pm 0.10$  &  9.5 & 4.0 & 11.0$-$12.4      &  4.426  &  3.959   &   4.749 &  4.268     \\ 
\enddata
\tablenotetext{a}{Vega isophotal wavelengths and IRAC zero magnitudes from Cohen (2005, priv. communication)} 
\tablenotetext{b}{Extinction from Indebetouw {\it et al.} (2005)} 
\tablenotetext{c}{\glimpse Point Source Catalog selection limits and brightness cutoff limits from Meade {\it et al.} (2005)}  
\tablenotetext{e}{Millions of sources} 
\tablenotetext{d}{``Effective'' Catalog sensitivity limit varies over the longitude range $|l|=10^{\circ}$ to $|l|=65^{\circ}$ }
\tablenotetext{f}{Southern Catalog/Archive is still missing $\sim$1\% of survey area} 
\end{deluxetable}




\end{document}